\documentclass[12pt, reqno]{amsart}
\usepackage{geometry}
\usepackage{amsthm}
\usepackage{amssymb}

\usepackage{enumerate}
\usepackage{graphicx}
\graphicspath{{./images/}}
\usepackage[T1]{fontenc}
\usepackage[utf8]{inputenc}
\numberwithin{equation}{section}
\usepackage[nomarkers]{endfloat}
\usepackage{hyperref}
\usepackage{cite}

\let\emptyset\varnothing

\title{ A Formal Analysis of Iterated TDD }

\author{Hemil Ruparel }
\address{ Pune }
\email{hemilruparel2002@gmail.com}
\thanks{Hemil Ruparel : 
Dedicated to my parents and family without their presence we are nothing. \\
}

\author{Nabarun Mondal}
\address{ Hyderabad }
\email{nabarun.mondal@gmail.com}
\thanks{Nabarun Mondal : 
Dedicated to my late professor Dr. Prashanta Kumar Nandi. \\ 
Dedicated to my parents. \\
In Memory of : Dhrubajyoti Ghosh. Dear Dhru, rest in peace.
}


\begin{document}


\begin{abstract} 
In this paper we  formally analyze the software methodology called (iterated) Test Driven Development (TDD).
We formally define Specification, Software, Testing, Equivalence Partitions, Coupling,
to argue about the nature of the software development in terms of TDD.
We formalize Iterative TDD and find a context in which iterated TDD ``provably produce'' ``provably correct code'' from ``specifications''
while being stable in terms of iterated code churns.
We demonstrate that outside this context iterated TDD will exhibit chaotic behavior, 
implying unpredictable messy amount of code churn. We argue that the research finding of ``ineffective'' iterated TDD
found by earlier researches are due to missing this context, while the findings of ``effective'' iterated TDD 
is due to accidentally falling into the context or simply placebo.

\end{abstract}

\maketitle

\begin{section}{Canonical Definition of Iterated TDD}

\begin{subsection}{Canon Definition}\label{canon-tdd}

We define TDD \cite{wtdd} as it is written in the canon article taken as the ``Definition of TDD''  \cite{ctdd} :

\begin{enumerate}
\item{
  Write a list of the test scenarios you want to cover}
\item{
  Turn exactly one item on the list into an actual, concrete, runnable
  test}
\item{
  Change the code to make the test (\& all previous tests) pass (adding
  items to the list as you discover them) }
\item{
  Optionally refactor to improve the implementation design}
\item{
  Until the list is empty, go back to [2].
}
\end{enumerate}
\end{subsection}

\begin{subsection}{Narrative}\label{nar-tdd}

The previous definition does not talk about any formal goals for iterative TDD.
Hence, we formalize the objective of TDD as follows:

To ensure that we end up having a
formally verifiable software in each step and in the end when all the
``scenarios'' are exhausted. Another ``optional'' objective is given as
to ``Improve the implementation design''. 

Note that it is not defined anywhere between
one implementation to other what can be ``improvement''.

This is not a good starting point to formally analyze the methodology,
as success metrics are not possible to be created on top of it. It is
very imprecise, and open to interpretations.

In this paper we propose a formal  methodology and provably
demonstrate how ``provably correct software'' can emerge with clear metric
of ``amount of code churn'' was done to attain it over the iterations -
albeit in a very narrow context. 

This practice we shall call formal Iterated TDD.
We are calling  the ``canon'' practice followed in the industry as ``Iterated
TDD'' for reasons which would be apparent shortly.

\end{subsection}

\end{section}

\begin{section}{Definitions}\label{definitions}

We would need some definitions to formalize the (Iterated) TDD
pseudo algorithm.

\begin{subsection}{Specification of Functions via Point
Pairs}\label{specification-of-functions-via-point-pairs}

Any function, computable or not, can be imagined to be pairs ( potentially $\aleph_1$ \cite{an} ) 
of input and output points in some abstract
space. It makes sense to describe functions by defining their specific
outputs at specific points or a large set of equivalent points. This
list of pairs we shall call point specification or ``specification'' for
brevity for the function it is trying to describe.

\begin{subsubsection}{Consistency}\label{consistency}

A function is formally defined as a relation where it is impossible to
have ``re-mapping'' e.g.~same input point mapped to two different output
points. The set of pair points must not have such spurious points, this
we shall call consistency criterion. This will become a key point in
case of software specification.

\end{subsubsection}

\begin{subsubsection}{Completeness}\label{completeness}

For functions which are well behaved this makes some sense. But even for
well behaved functions this is not a good enough approximation.

Take a nice function like $f(x) = x$ , identify function, but one can not
define this function by keeping on adding pairs of specification values. 

A much more interesting function like $f(x) = sin(x)$ is much harder to
describe, although we can always define them pointwise, and that would
ensure the resulting ``sampling'' looks much and much like the target
function, one must understand infinite pairs would be required to
specify $sin(x)$. Even with $\aleph_0$ points specified, there would
be set of infinite family of functions who are not $sin(x)$ but just
gives off the exact same value at all those specific points. This has a
name, called pointwise convergence \cite{pc}.

Outside those fixed set of points the family of functions can take
arbitrary values,  and thus specification via point pairs arguably pose
a problem.

Luckily, for software we can do much better, which is the topic for the
next section.

\end{subsubsection}

\end{subsection}

\begin{subsection}{Software}\label{software}

A software is defined to be a Computable Function - mapping abstract
vector space of input to the output vector space. The notion of using
vectors is due to all real software works with many inputs and hence the
state space is multidimensional which is the exact same space as output.

$$
S : \hat I \to \hat O
$$

where $\hat I := <x_i>$ is the input vector while $\hat O := <y_j>$ 
is the output vector. These vectors are defined not
in physics sense, but pure mathematical sense. The only change between
the pointwise defined function vs specified software is about being
``Computable'' \cite{comp}.

\end{subsection}

\begin{subsection}{Software Test}\label{software-test}

A  ``Software'' test is defined as a higher order function \cite{hof} : 
$$
T : t  < \hat I_t, S_t , \hat O_e > \to ( S_t( \hat I_t) := \hat O_t)= \hat O_e
$$

In plain English, a test is comprise of Input vector $\hat I_t$,
the software under test $S_t$, and the expected output vector
$\hat O_t$ , it runs the $S_t$ with the input, and checks whether or
not the expected output $\hat O_t$ matches against the actual output
of the system $S_t( \hat I_t) := \hat O_t$, and it simply checks
whether or not $\hat O_t = \hat O_e$ , hence the range of the test is
Boolean.

A software test, then contains a single point specification for the
desired Software, this is the test vector \cite{tv}.

A software test does not need to be computable in general.
Unfortunately, any automated test, by definition needs to be computable.
This also pose a problem for testing in general. Example of a test that
is not computable \cite{dec} \cite{hlt} can be : a human reporting software has hung or went
into infinite loop. This is impossible to do algorithmically, unless we
bound the time. This sort of scenarios comes under Oracles in
computation \cite{om}.

\end{subsection}

\begin{subsection}{Code : Control Flow Graph, Branches}\label{cfg-branches}

Software is written essentially using arithmetic logic and then
conditional jump - this being the very definition of Turing Complete 
languages \cite{tcomp}. This structure with conditional jump ensures that the
different inputs takes different code paths. A code path is a path (
even having cycle ) in the control flow graph  \cite{cfg} (CFG) of the software which
starts at the top layer of the directed graph that is the code and ends
in the output or bottom later.

Formally we can always create a single input node and output node in any
control flow graph.

Treating multiple iterations of the same cycle as a single cycle, we can
evidently say given the nodes of the graph is finite, there would be
finite (but incredibly high) number of flow paths in the graph.

\end{subsection}

\begin{subsection}{Partitions : Equivalence Classes}\label{equivalence-class}

At this point we introduce the notion of equivalence class of input
vectors to software. If two inputs $\hat I_x$ and $\hat I_y$ takes
the same path $P$ in the control flow graph, then they are equivalent.

This has immense implication in testing and finding tests. Because this
induces an equivalence partitioning on the input space itself, because
all $\hat I_x$ in the same equivalence class can be treated as exactly
equivalent, because all of them would follow the exact same code path \cite{gpath} in
the control flow graph. There is another related concept called boundary
value analysis (BVA) \cite{bva}, but we would not go there, because that is not
going to alter the subsequent analysis in any significant way.

This effectively means by isolating all equivalence partitions and
choosing one input member from each of them we can test the system the
most optimal way - by restricting the number of ``Software Test''s, as well 
as providing a full ``coverage'' in terms of specification.

For example, if there are $A,B,C,D$ equivalent
classes \cite{eqc},  then choosing $\hat I_A \in A$, only one would test the code
path for $A$, similarly for the rest. So instead of infinite inputs,
only 4 inputs would suffice. Notice that these are the most optimal set
of inputs, the bare minimum to ensure that the system works in a
provably correct manner.

This formally brings the problem to finding the exhaustive set of equivalent classes 
( let's call it  $\mathbb{E}$ ) that completely describes one implementation of a ``Software'' system.

That is impossible without the implementation. It is wrong to perceive
that this technique is driven by specification alone. EQCP is a gray box
testing  \cite{gbt} technique as it requires assuming some implementation details\cite{eqcp}.

What would be an upper bound of the number of such equivalent classes ?
This depends on the number of the conditional jumps. It is easy to prove that if there are
$B$ branches, then the  bound for the number of the equivalence class is $O( 2^B )$
where $O(.)$ is ``Big-Oh'' one of the Bachmann Landau asymptotic notations \cite{bigo},
This also would be very important for a pragmatic discussion later.

The Equivalent classes would be called EQCP from now on because they
partition the input set into Equivalent Classes. There would be many
EQCP for individual ``features'' in ``Software''.

\end{subsection}

\begin{subsection}{Coupling in Software}\label{coupling-in-software}

At this point we introduce the phenomenon of coupling \cite{coup} between Equivalent
Classes, when seen with respect to code implementation.

Given individual EQCP are depicting unique paths in the control flow
graph (CFG), then coupling said to exists between EQCPs $E_x$ with
path $P_x$ and $E_y$ with path $P_y$  if and only if 
$P_x \cap P_y \ne \emptyset $.

That is, if paths \cite{gpath} $P_x, P_y $ has some common nodes, then $E_x,E_y$
are coupled. In fact we can define the amount of coupling using
similarity measures now, most easy one would be a Jaccard distance \cite{jac} like
measure: 

\begin{equation}\label{coupling}
C(E_x,E_y) = \frac{ |P_x \cap P_y| }{ |P_x \cup P_y|  }
\end{equation}

This essentially says - ``Measure of the coupling between two equivalent
classes is the amount of code shared between them relative to all the
unique code path they have together''. We need to understand that even
code shared for good reason, like applying DRY \cite{dry} and not doing it even
methodically also would create coupling via this definition. Any shared
function between two EQCP would mean coupling exists. As we shall see
Coupling  becomes a key phenomenon while analyzing the stability of
software under Iterative TDD.

\end{subsection}

\begin{subsection}{Test Driven Development as Equivalent Class
Specification}\label{test-driven-development-as-equivalent-class-specification}

We can now formally define a software system specification in a finite,
and provably correct way.

If we can just specify the equivalence classes, then we can just fix
the software output at those specification points and the resulting tests
precisely, and correctly defines the software behavior. This must be
taken as the formal definition of (non iterative, formal) TDD with
absolute minimal test inputs:

\begin{quote}
Given an abstract (not written) Software $S_a$, let's imagine the
equivalence classes $E_x$ such that $E_x, E_y$ are independent and
specify the input and output expected from each equivalence classes.
Now, ensure all of these tests pass by writing the implementation.
\end{quote}

This system is provably complete and correct, by construction. Every
test just ensures all individual EQCP behavior is passed via
construction. Given that was the entire specification, this means the
system passes all criterion for the specification, and thus becomes
provably correct.

The input output specifications can be immediately translated into
tests, and that gives the formal provable meaning to TDD. Any random
tests on features won't do, it have to be (at bare minimum) spanning
the entire EQCP (the formal specification points).

This is the real superpower of TDD, formal verification baked into
development. Although, truth to be told, this way of constructing
software has been known for many decades.

And this is why the canonical TDD  was called out as ``iterated TDD'' 
because this formal non iterative TDD model does not include change of specification,
thereby does not follow any iteration and thus does not consider code churn thereof.
This formal non iterated model is one single shot transformation of bunch of specifications points into code 
via transforming them into EQCP.

\end{subsection}

\begin{subsection}{Practical Correctness of TDD}\label{correctness-of-tdd}

The correctness of TDD for a practical application hinges on the
following :

\begin{enumerate}
\item{
  Is the specification complete enough ( to take care of all the
  equivalent classes )? 
 }
\item{
  Is the specification non contradictory ?
}
\end{enumerate}

That it is impossible to get (1,2) done together follows from Godel's
Incompleteness theorems \cite{inc}, but that is applicable to any specification,
not only Software. Thus this argument should not be admissible as
failure of TDD in itself.

Now we ignore the notion of contradiction and focus on completeness and
stability when one tests gets added at one time ( iterated or incrementally changed specification TDD ).

\end{subsection}

\begin{subsection}{Practical Completeness of TDD Spec}\label{completeness-of-tdd-spec}

The business specification should be such that the formal specification
of all possible Equivalence classes must be drawn from it. As it is
bounded by $O( 2^B )$ - this itself is not remotely possible. To
understand how this bound works, a simple program unix \texttt{cat} has
more than 60 branches \cite{cat}. The equivalent class specification of this
program is bounded by $2^{60}$ and the total stars in the universe
are estimated to be $2\times 10^{24}$ for comparison.

But this huge numbers  does not disprove the crux of TDD, it only points to the
fact that formal EQCP is a practical challenge and to be handled
pragmatically, probably via reducing the specification scope further and further.

\end{subsection}

\end{section}

\begin{section}{Analysis of Iterated TDD}\label{analysis-of-iterated-tdd}

\begin{subsection}{Development under TDD}\label{development-under-tdd}

Note that the methodology does not specify how to implement the paths of
each equivalent classes in the code. Hence evidently there is no way it
can ever improve on the ``non correct aspect of quality'' of software,
one of them would be to lower coupling. In fact if not controlled this
would bring in way more coupling than it was required due to application
of other principles like DRY. Because there are infinite way to conform
to the ``point wise convergence'' but then the methodology does not
specify any family of approach to do so. These are some of the key open
problems of the methodology as it formally stands as of now.

A trivial  non coupled way to construct code would be such that  no equivalence class
share any code path. This would solve the coupling problem, but code
would be massively bloated. Any other way would reduce the code but
ensure the classes would be coupled to some extent.

This is a choice. We want to simultaneously minimize two metrics:

\begin{equation}\label{min-couple}
C_S = \sum\limits_{x \ne y} C(E_x,E_y)
\end{equation}

along with:

\begin{equation}\label{min-couple}
S_S = min _{n} \{  K(S_n) \}
\end{equation}

where $S_S$ stands  for ``source code size'' where $K(S_n)$ defines 
the  optimal code size of the System  $S$ at  $n$'th implementation trial.
This is a very hard problem as Chaitin Solomonoff Kolmogorov Complexity (CSK)  \cite{kc} is \textbf{non Computable} \cite{comp}.
We do not even know if such a problem can be solved in formal setting.
We posit it as an open problem in Software Development.

In lieu of that we continue in our analysis 
where we imagine a bit of necessary code coupling and try to reduce the code churn in terms of EQCPs.
This coupling would have implication in iterated TDD, and we show a provable
methodology that can reduce code churn in the later sections.

\end{subsection}

\begin{subsection}{Iterated TDD}\label{iterated-tdd}

An Iterated ( incremental) TDD is when we add more specification to the mix of
already existing ones one step at a time under practical setting. 
This \textbf{incrementally added test based iterative TDD methodology} is 
what we discuss in the next sections as this is the one which proponents of
TDD talks about. We note down it is different from the formal TDD 
we have established before - canon TDD is an iterated version of the formal TDD
with specifications being added per iteration.

\end{subsection}

\begin{subsection}{Stability of EQCP under Iterated TDD}\label{stability-of-eqcp-under-iterated-tdd}

Suppose, there is already an existing system in place with tests done
the right way - following the EQCP method discussed earlier, e.g. following TDD.

Is it possible to add more specification w/o rewriting existing
equivalent classes in a stable manner? 

The sort of stability we are looking for is called BIBO Bounded input Bounded Output stability \cite{bibo}, 
that is, for a small change in specification, not much change would happen in the EQCP space.

This is the iterative TDD, applying this again and again.
The answer to this is key to the prospect of iterative TDD.

Formally, Software $S_r$, has the equivalent classes $E_x \in \mathbb{E}_r$ , 
and now more specification augmentation is happening. 
The following questions need to be asked:

\begin{enumerate}
\item{
  How many of the existing EQCP will not be effected by this?
}
\item{
  How many new EQCP needs to be added?
}
\item{
  How many EQCP needs to be removed?
}
\end{enumerate}

As one can surmise, this is the transformation step of a fixed point
iteration on the abstract space of the EQCP. We shall get back to it
slightly later. 
\end{subsection}

\begin{subsection}{Additional Branching}\label{additional-branching}

The answer to the question [2] is in isolation if there would be $K$ branches to implement the delta specification 
- new feature then, the isolated equivalent classes would be in $O(2^K)$ , thus, 
the minimum new classes needed would be bounded by this value.

At most it can impact every equivalence class and at least it adds
$O(2^K)$ classes and hence tests. So, at the best case scenario, the
total branches would become $O(2^B + 2^K) = O(2^B)$ given $B >>  K$. 
The complexity increases, but not drastically, unless $B = O(K)$.

\end{subsection}

\begin{subsection}{Impact of Coupling}\label{impact-of-coupling}

What happens when there is coupling? Instead of adding the terms, now
because of dependency, the terms gets multiplied. Thus, with coupling
the resulting complexity becomes $O(2^B \times 2^K) = O(2^{B+K})$ .
The delta change results in exponential growth even if $B \ne O(K)$.

This is a problem.

If the implementation of those equivalent class was such a way that
there was minimal coupling, then less classes would be impacted via this
step in the iteration. But this is not a principle of TDD in the first
place in any form in any practical application of software development.
In fact software principle like DRY and modular programming would mandate code sharing, 
and hence there would always be some coupling.

\end{subsection}

\begin{subsection}{Iterated TDD as a Dynamical System}\label{iterated-tdd-as-a-dynamical-system}

At this point we can formally represent iterated TDD as a dynamical
system \cite{dyn}.

As discussed, this EQCP merging culminates into a lot of those
equivalence classes being thrown out, new classes being created - a
fixed point iteration on the abstract space of the EQCP itself, which we
can now formally define as follows: 

\begin{equation}\label{iter-tdd-eqcp}
\mathbb{E}_{n+1} = \tau( \mathbb{E}_n , \delta_n)
\end{equation}

Where at step $n$,  $\mathbb{E}_n$ is the current set of EQCPs,
while based on new specification ( $\delta_n$ ) and the $\mathbb{E}_n$ TDD system $\tau$ 
produces new set of EQCPs ( $\mathbb{E}_{n+1}$ ) for the next step $n+1$. 

This is the fixed point iteration of incremental software development from point pair specification or incremental, iterated TDD.

It is obvious that the first ever specification was done with empty
equivalent classes ( $\mathbb{E}_0 = \emptyset$ ) and initial specification of $\delta_0$: 

$$
\mathbb{E}_{1} = \tau( \emptyset, \delta_0 )
$$

This is how formally iterated or incremental TDD looks like.
These equations now depicts a dynamical, complex system with am initial boundary value or starting condition.

\end{subsection}

\begin{subsection}{Stability Space}\label{stability-space}

While EQCP space is nice to visualize what is happening for real in
terms of Software Specification and Test cases, it is not descriptive
enough to translate into numbers so that we can track the trajectory of
the Dynamical System. 

How much change in the EQCP space is happening on each iteration of iterated TDD?
It is impossible to comprehend that in the EQCP space.

For gaining this insight we would need a metric, that would define
how stable the system is over the iterations in terms of retaining past
EQCPs - how much code remained same between iterations.

We define the stability metric as follows :

\begin{equation}\label{iter-tdd-stability}
\Sigma_{n+1} = 1 - \frac{|\mathbb{E}_n \cap \mathbb{E}_{n+1}|}{|\mathbb{E}_n \cup \mathbb{E}_{n+1}| } \; ; \; \Sigma_n \in \mathbb{Q} \cap(0,1)
\end{equation}

The stability metric $\Sigma$ also depicts a metric space \cite{ms} with
distance between two stability points $a,b \in \Sigma$ as defined to be : $ d(a,b) = |a-b| $.

\begin{subsubsection}{Stable Point : 0 }\label{zero-stable}

Observe the following, if we ensure that no EQCP has any shared code,
then the only way to make change is to simply add new code, and thus
$\mathbb{E}_n \subset \mathbb{E}_{n+1}$, 
and that gives minimum value of $\Sigma$ if and only if  
$| \mathbb{E}_{n+1}  \setminus  \mathbb{E}_n| $ 
can be minimized .

\textbf{A value of $\Sigma$ close to 0 shows the system has been very
stable between last to the current iteration.} This is when ``very
loose'' coupling ensured that we can create branches which do not
interact with existing branches that much. We present order of magnitude
estimates for ``highly stable'' uncoupled $^U\Sigma$ value as follows:

\begin{equation}\label{iter-tdd-st-0}
^U\Sigma_{n+1} \approx 1 - \frac{|\mathbb{E}_n| }{|\mathbb{E}_{n+1}| } \approx  1  - \frac{O(2^B)}{O(2^B + 2^K)} \approx 1 - \frac{1}{1 + 2^{K-B}} \approx 0 \; ; \; B >> K
\end{equation}

We note that it is impossible to reach value 0 under any circumstances
other than when $\mathbb{E}_n = \mathbb{E}_{n+1}$ which means, the
specification \(\delta_n\) did not change anything in EQCP space, e.g.~a
complete dud or spurious specification.

Importantly, there can be cases where even without coupling, as demonstrated by :
$|\mathbb{E}_n | << | \mathbb{E}_{n+1}|$ , then even though
$\mathbb{E}_n \subset \mathbb{E}_{n+1}$, the stability would be going
for a toss - this is driven by having $B = O(K)$.  

\end{subsubsection}

\begin{subsubsection}{Unstable Point : 1}\label{unstable-one}

Now the other side of the coin is when
$\mathbb{E}_n \cap \mathbb{E}_{n+1} \approx \emptyset$, in this case the
value of $\Sigma$ goes to 1.

\textbf{A value of $\Sigma$ close to 1 shows the system has been very
unstable between last to the current iteration.} This is when ``strong''
coupling ensured that we need to rewrite a lot of the EQCP
implementations in code.

The ``reasonably coupled''  $^C\Sigma$ estimate would be as follows: 

\begin{equation}\label{iter-tdd-st-1}
^C\Sigma_{n+1} \approx 1 - \frac{O(|\mathbb{E}_n \cap \mathbb{E}_{n+1} |) }{O(| \mathbb{E}_n \cup \mathbb{E}_{n+1}|) } \approx  1  - \frac{O(2^B)}{O(2^{B + K})} \approx 1 - \frac{1}{2^K} \approx 1 \; ; \; K >> 1
\end{equation}

Where $K$ is some constant estimating the branch changes due to $\delta$ as depicted in previous section.

\end{subsubsection}

\end{subsection}

\begin{subsection}{Guiding Stability Algorithm}\label{guiding-stability-algorithm}

Assuming coupling would almost always be present, one way for us to
avoid unpredictable jumps in the stability, we can device our
development strategy such that the $\Sigma$ does not change
drastically towards 1. 

At this point, if there were many alternative way
to program ( $P_i$ ) the $\delta_n$ change, we may want to chose the
alternative \(P_x\) way to program which minimizes $\Sigma_{n+1}$. 
If we do, then the system remains stable in the short term. But this is a
direct anti thesis  of ``less code change and faster changing ability'', as it minimizing
$\Sigma_{n+1}$ culminate into more code change, because it would
inherently try to lose some coupling!

More importantly, this computation of minimizing the $\Sigma$ 
post applying the $\delta$ change can be greedy, 
but it is evident that here is where hill climbing creeps up,  there can be a minima hidden somewhere else.

At this point, in the worst case it would boil down to applying all specification changes $ \{ \delta_i  \}$
which would have have a factorial runtime or, would be in NP.
This is anti agile, and definitely not ``small incremental change'', this is a lot of change, pre-computed, 
and applied to minimize code churn.

By this time, we have understood that practically following  guided stability is already very hard, however, worse is yet to be seen by us.
Unfortunately even with this guided approach there would be some problems which would not go away, 
in the long term, that is the discussion of the next section.

\end{subsection}

\begin{subsection}{Chaos in Stability space}\label{chaos-in-stability-space}

We now proceed to demonstrate that the iteration driven by $(\tau,  \delta_n)$
in Stability Space $\Sigma$ has characteristics of a system capable of
showcasing chaotic dynamical behavior \cite{wchaos}.

Given there is no universally agreed definition of chaos - we - like
most people would accept the following working definition \cite{chaos1} \cite{chaos2}:

\begin{quote}
Chaos is aperiodic time-asymptotic behavior in a deterministic system
which exhibits sensitive dependence on initial conditions.
\end{quote}

These characteristics would now be demonstrated for iterated TDD.

\begin{enumerate}
\item{
  \textbf{\emph{Aperiodic time-asymptotic behavior}} : this implies the
  existence of phase-space trajectories which do not settle down to
  fixed points or periodic orbits. For practical reasons, we insist that
  these trajectories are not too rare. We also require the trajectories
  to be \emph{bounded}: \emph{i.e.}, they should not go off to infinity.

  The sequence $\Sigma_n \in \mathbb{Q} \cap (0,1)$ is bounded by
  definition. The trajectories are not rare, and it is practically
  impossible for the sequence to settle down to periodic orbits or
  converging sequence. Note that w/o the presence of coupling this
  sequence can be made to orbit around approximating 0 most of the time.
}

\item{
  \textbf{\emph{Deterministic}} : this implies that the equations of
  motion of the system possess no random inputs. In other words, the
  irregular behavior of the system arises from non-linear dynamics and
  not from noisy driving forces.
  
  One can argue that the sequence is driven by $\delta_n$ -
an external input, but it is not. Iterative TDD has this baked in, as
part of the system iteration description , and the processing of it is
algorithmic in the formal methodology which we present for formal
correctness for the software. In fact we can argue that the sequence
$\delta_n$ can be specified beforehand, and it would make it fully
deterministic and it would not impact our analysis.

}

\item{
  \textbf{\emph{Sensitive dependence on initial conditions}} : this
  implies that nearby points can be spread further over time while
  distant points can come close over time - e.g.~stretching and folding
  of the space. In fact it is said to be:

  \begin{quote}
  Chaos can be understood as a dynamical process in which microscopic
  information hidden in the details of a system's state is dug out and
  expanded to a macroscopically visible scale (\emph{stretching}), while
  the macroscopic information visible in the current system's state is
  continuously discarded (\emph{folding}). The system has a
  positive Lyapunov exponent \cite{le}.
  \end{quote}

  This is evident in case of coupling.

  CFG comprise of the micro details which culminates into the the space
  of EQCP, and merging further specification over that produce the
  sequence $\Sigma_n$. Inherently a lot of micro details are being
  pushed into visibility and then again being discarded as in the
  $\Sigma$ space, the information about current complexity of the
  system ( EQCP space $\mathbb{E}$ )  does not exist.

  We shall now proceed to formally demonstrate that Lyapunov exponent is positive for $\Sigma$.

  Given two nearby points in $\Sigma_n$ , say $a,b : |a-b| < \epsilon $ , 
  there is no guarantee that in next iteration how further apart the sequence would go, given even exactly
  same specification of $\delta_n$ . Let $\Sigma(p, \delta)$ be the
  next iteration sequence after starting from $p$ in $\Sigma_n$ post
  applying the same specification change  $\delta$.

  Then  
  $\textbar{}\Sigma(a, \delta) - \Sigma(b, \delta)\textbar{} \ne 0 $ 
  holds true almost always for all practical purposes.

  Let us define the function $\Delta(a,b, \delta )$ as follows:

  \begin{equation}\label{st-diff}
  \Delta(a,b, \delta) = \frac{|\Sigma(a,\delta) - \Sigma(b, \delta)|}{|a-b|}
  \end{equation}

Then, a \textbf{stretch} happens when $\Delta(x,y, \delta) > 1$ and a
\textbf{fold} happens when $\Delta(x,y, \delta) < 1$. 

This is to say, stretch increases the distance between the trajectories starting with
$(a,b)$ while fold reduces it. We notice that the definition of Lyapunov exponent of
the $\Sigma$ would be as follows:

\begin{equation}\label{st-lp}
\lambda = ln(\Delta(a,b,\delta))
\end{equation}

We can approximate $\Sigma(x,\delta)$ in presence of some coupling -
where $B_x$ is the branching at $x$ and $K_x$ is the addition of
branching due to application of $\delta$ as follows ( estimating from previous section): 

$$
^C\Sigma(x,\delta)  \approx 1 - \frac{O(2^{B_x})}{O(2^{B_x + K_x})} \approx 1 - \frac{1}{2^{K_x}}
$$

This when substituted reduces to:

$$
\Delta(a,b,\delta) \approx  \frac{ |\frac{1}{2^{K_a}} - \frac{1}{2^{K_b}} | } {|a-b|} \approx \frac{ |2^{K_a} - 2^{K_b} | } {2^{K_a+K_b}|a-b|}
$$

 Now we choose a suitable $\epsilon$ for our purpose to simplify the expression as well as minimize it: 

$$
\epsilon <  \frac{1}{2^{K_a+K_b}}
$$

Thus making the smallest bound possible for $\Delta$ as : 

$$
\Delta(a,b,\delta) \approx |2^{K_a} - 2^{K_b} | \approx \theta(2^L) \; ; \; \forall (K_a \ne K_b) \; L > 1
$$

And this immediately demonstrates that 
Lyapunov Exponent for the system is positive ( $\lambda > 0$ ) : 

\begin{equation}\label{st-lp-final}
\lambda = ln(\Delta(a,b,\delta)) \approx L \times ln(2)  \; ; \; \forall (K_a \ne K_b) \; L > 1
\end{equation}

thereby proving that the $\Sigma$ map is expansive and hence
Chaotic under the influence of coupling.
}
\end{enumerate}

We can argue the same in a semi formal way.

Evidently, if only folding happens, then every sequence would converge.
This is an extreme view. In the same way if only stretching happens,
then because the sequence is bound, it must converge again to 0 or 1.
This is another extreme view.

We can safely say the probability that for every tuple $(a,b, \delta)$
that the $\lambda > 1$ would be $0$. So goes the same for $\lambda < 1$.

It is much more plausible that a function like this would have some
intervals where it would stretch and some intervals where it would fold
depends on the $\delta$. This is the most likely phenomenon which
invariably would generate a sequences diverging and converging in
$\Sigma$ thereby producing the dynamic process that stretches and
folds - and thus creating sensitive dependence on initial condition, the
hallmark of chaos.

The above points make it very clear that the sequence $\Sigma$ may
show all properties of chaotic dynamics. Which proves that iteration of
iterated TDD can and would show chaotic dynamics.

\end{subsection}

\end{section}

\begin{section}{Practical Considerations for Software Development Under Iterative TDD}\label{practical-considerations-for-software-development-under-iterative-tdd}

\begin{subsection}{Identifying Chaotic Trajectory}\label{identifying-chaotic-trajectory}

Is there a guarantee that chaotic patterns would emerge on each case? No
one knows. Chaos in software development \cite{sdchaos} has been discussed about 
although not in much formal details like this. If we are very lucky it
would not, but it is hard to tell. Only by carefully monitoring the
sequences we would be able to claim whether we entered any chaotic
sequence or not and this formalism gives a metric such that the sequence
can be tested for emergence of chaos - by following Kantz  \cite{KANTZ199477}. 
That would be the empirical way of measuring on each iteration how the
progress is happening. Given agility is the name of the game now, we can
add 52 data points a year for each project if weekly shipping of
software is followed.

\end{subsection}

\begin{subsection}{Domain of Stability for Iterated TDD}\label{domain-of-stability-for-iterated-tdd}

Let's imagine the worst case, almost all of the sequences would be
chaotic.

What is so problematic about chaotic dynamics appearing in the phase of
``stability'' of EQCP ? This means there might a unpredictable amount of churn in terms of
the changes in the EQCP. And that means churns in the ``pair
points specifications'' e.g tests which were to ``hold the correctness
of the software'', implying a unpredictable, possibly a very high implementation change MUST happen.

If in one iteration which was created by a tiny change in specification
impacted 50\% of the test cases to refactor source code and tests
thoroughly, evidently this would become a huge problem.

The chaotic thesis suggests that not this is only possible, but also
highly likely due to the mixing of EQCPs in terms of coupling, and a
direct result of code refactoring trying to apply DRY principle.

Hence the formal idea of just fixing input output points and rapid, small iteration on specification 
can not work in general unless we keep on reducing the scope of the specification. 

It is only guaranteed to work (produce provably correct software and predictable amount of code churn) 
at the lowest abstraction level if there are very less coupling by definition.
Unfortunately the proponents of TDD want to make it work even at user
specification level - where it entirely lose out  its rigor and has no
provable applicability to either improve the quality of the product or
the code itself.

\end{subsection}

\begin{subsection}{Uncertainty Principle of Iterated TDD}\label{up-iter-tdd}

We have uncovered an uncertainty principle \cite{fup} of sorts here:

\begin{quote}
With coupling at play, if we try to fix more specification by specifying more EQCP, 
then the code churn becomes unpredictable. And if we do not go exhaustive on EQCP, then the 
formal correctness software producing characteristics of the methodology disappears.
\end{quote}

It seems in the presence of coupling, we can 
either choose formal correctness or 
choose code churn stability, 
not both.

This insight is unheard of, but the theory points us in this direction. 
If the chaotic thesis is correct, this is to be taken as a foundational law of Software Engineering.

While this demonstrates why coupling is a problem, however, this is much stronger thesis, 
this tantamount to \textbf{any shared code is a problem if the code supposed to change later}.
\end{subsection}

\begin{subsection}{Revisiting Guided Approach}\label{guided-revisited}
Readers may argue that how then this analysis does not apply to any other software development process?
The answer lies in the guided approach. In case, if one does not make the software fixed via hard test driven specification, 
then there is loss of ``correctness'' - granted, but there is a lot of ``wiggle'' room to build the system.

With the guided approach  one can even try to avoid the entire chaotic trajectories by prioritizing specifications
or even rejecting it for the time being, till a suitable time comes to apply such that the stability is not changed that much.

This, evidently is what non agile waterfall, or iterated waterfall  \cite{idm}was all about. 
In fact we are formally defining prototypical development at this point \cite{pdm}.

Would they avoid the unstable paths? Sometimes. But mostly they would make the system ``slower'' in the stability space.
Here, we are not talking about the slowness of delivery, we are talking about slow movement of the system in the stability space.
This way, it would take a very long time to reach a chaotic state.

\end{subsection}

\begin{subsection}{Path Forward - Approaches}\label{path-forward---approaches}

From the last section to avoid these chaotic sequences we can try
avoiding all of these by either:

\begin{enumerate}
\item{
  Making the specification more relaxed - at that point it would specify
  almost nothing and there would be almost no chaotic behavior because
  of the state space of EQCP being reduced drastically. 
  This is the a cargo cult approach, producing only placebo, the application of TDD w/o any formalism.
}
\item{
  Or, we can try to decrease coupling, in which case it would bloat the
  software by not having shared code path - this would result is
  unimaginable bloat in the software - given we are looking at very
  large dimension of EQCP state space.
}
\end{enumerate}

Evidently, then via [2] iterated TDD, therefore, can only be
effectively done in practice when the $\mathbb{E}_n$ space is
extremely small and the context of ``Software'' is very narrow.

\end{subsection}

\begin{subsection}{Context Of Applicability}\label{context-of-applicability}

Not all is lost however. As it is proven, if we can go narrower and
narrower, to the point when EQCPs stop effectively sharing code with one another,
TDD  becomes formally correct,  also the  methodology
to develop software in regular iteration with predictable churn. This narrow specification
contexts are in fact  the unit tests with very less coupling 
which  guarantee of becoming chaos free! 

We can now formally define scope for formal iterated TDD, which is guaranteed to
work - e.g.~create formal verifiable correct software as follows without
ever destabilizing source code:

\begin{quote}
Unit like tests where implementation of such features do not share any
source code, e.g. Independent (completely decoupled) - such that in
every iteration the decoupling holds true guarantee to hold to
verifiable correct behavior.
\end{quote}

And it is in this context TDD reigns supreme. Anything other than that -
correctness or stability can not be guaranteed. Just like one can try to
use a scalpel to dig a canal, it just won't work. Any effort of using
the scalpel to create a canal is not only misguided, but futile, and not
even wrong.

Do iterative TDD, just ensure all EQCPs are completely decoupled, this, 
now becomes a formally correct software producing code churn wise stable methodology. 
Now, in practice it is hard to do, even for Unit tests, so a small amount coupling should not really
harm the effectiveness via that much - but at that point Chaotic behavior stems in.

Principles like AHA, WET \cite{dry} comes in extremely handy in this regard. Even
with very less coupling there is no absolute guarantee of code
stability, due to emergence of chaos but at least we are in the right
track by being formally correct, and the resulting chaos can be tamed.

\end{subsection}

\begin{subsection}{A Perspective on Popular ``Business Specification based'' TDD}\label{a-view-of-popular-business-specification-based-tdd}

The previous issues culminates into less and less specific
specifications used in the industry. At that point they cover so less
equivalence classes that TDD would lose all it's effectiveness which is
to be found rigorously at the unit test level. Thus we do have a
problem, if we specify more and more, the resulting software has high
coupling thereby ensuring the iterations are destabilized. If we specify
less and less the resulting diluted TDD is just homeopathy, water in the
name of medicine but peoples believe making it ``work'' - a placebo \cite{Bakhtiary2020-yg}.

This is not hard to understand, as TDD mandates writing the tests first, there are some tests, for sure, 
better than none, and this essentially ensures there is at least some correctness in the mix.
The fear of failing tests ensures code is often correctly written.
It has been well understood that developers tend to write better code just because there would be testers who would test it.
This however does not consider the ``cost'' of stability in code churn. This metric, surprisingly was never studied!

Interestingly ``Business Specification Driven TDD''  is the most popular TDD in the industry. 
This ``Some input,output are verified'' is not really an effective
methodology, given the nature of the number of tests required runs in
exponential numbers in terms of the EQCP for the features.

However, it gives a lot of people something to talk about and mental peace just like Homeopathy
sans effectiveness other than placebo as it was found out in another research : \cite{tdd_3}.

We can also safely say, any low level, low coupled, EQCP based formal TDD method would be reasonably successful, 
if those practices were to be followed, iterated TDD would definitely be very effective.
There are some publications where it has been shown to do exactly that \cite{Bakhtiary2020-yg}.

\end{subsection}

\begin{subsection}{Cargo Cult ``Software'' Engineering? }\label{cargo-cult}

We can therefore conclude that iterated TDD without understanding the applicability context
is like washing your hand with water before you eat, 
while the ``washing hand'' would be a good practice, but if the water used was filthy, it would degenerate to numerable problems.
This is the status of industry with respect to TDD, for those who are into the right context, it works, give or take.
Those who are not, it does not.

We conclude by making a much more starker remark, the proponents of TDD, or ``industry best practices''
stopped asking ``is this effective or provable''  a long time ago. 
Their new established position is : ``No evidence required for common sense practices''.
In fact, this is the verbatim response when asked about  efficacy and provability of some of the best practices:

\begin{quote}
You want to debate seriously? Then you have to drop the ridiculous sense that ``Good Practices'' require 
scientific evidence before they can be realized to work - which would disprove much of the ``Good Practices''
which are ``successfully used'' in the industry.
\end{quote}

Even if we ignore the irony of the previous quote, one but just wonder if evidently Software had become entirely cargo cult \cite{ccsc}, the above quote proves it beyond doubt.
Very few admit it openly, but it is what it has become.

\end{subsection}

\end{section}

\begin{section}{Closing Remarks}\label{closing-remarks}

Formal iterated TDD, as presented here, is shown to produce correct
software code. The issue with such production requires a lot more formal
and practical considerations.

When done correctly (by EQCP and reducing coupling between them) it ensures we can further add
more features to the existing software while maintaining stability as
well as correctness as we go.

If that reduction of coupling is not followed, then the addition of more
equivalence classes could and most definitely would modify a significant
amount EQCP mapping by ensuring one must rewrite a very significant
amount of tests, as well as implementations. This is also seen in
reality. Anything at any further higher level of abstraction that Unit
like tests would have impact like placebo.

Hence we propose iterated TDD is to be done at the Unit Testing level
only, where it works correctly and satisfactorily because of Units
should be essentially maximally decoupled keeping an constant eye on the
coupling generated by those tests being constantly added, which is hard,
but not impossible to do and shows provable theoretical efficacy:
provably correct software production along with predictable code churn.

\end{section}

\bibliography{tdd}{}

\begin{thebibliography}{10}

\bibitem{wtdd}
Various, ``{Test Driven Development}.''
  \url{https://en.wikipedia.org/wiki/Test-driven_development}, 2024.
\newblock [Online; accessed 3-July-2024].

\bibitem{ctdd}
K.~Beck, ``{Canon TDD}.'' \url{https://tidyfirst.substack.com/p/canon-tdd},
  2023.
\newblock [Online; accessed 3-July-2024].

\bibitem{an}
``{Aleph Numbers}.'' \url{https://en.wikipedia.org/wiki/Aleph_number}, 2024.
\newblock [Online; accessed 3-July-2024].

\bibitem{pc}
``{Pointwise Convergence}.''
  \url{https://en.wikipedia.org/wiki/Pointwise_convergence}, 2024.
\newblock [Online; accessed 3-July-2024].

\bibitem{comp}
``{Computability}.'' \url{https://en.wikipedia.org/wiki/Computability}, 2024.
\newblock [Online; accessed 3-July-2024].

\bibitem{hof}
``{Higher Order Function}.''
  \url{https://en.wikipedia.org/wiki/Higher-order_function}, 2024.
\newblock [Online; accessed 3-July-2024].

\bibitem{tv}
``{Test Vector}.'' \url{https://en.wikipedia.org/wiki/Test_vector}, 2024.
\newblock [Online; accessed 3-July-2024].

\bibitem{dec}
``{Decidability in Logic}.''
  \url{https://en.wikipedia.org/wiki/Decidability_(logic)}, 2024.
\newblock [Online; accessed 3-July-2024].

\bibitem{hlt}
``{Halting Problem}.'' \url{https://en.wikipedia.org/wiki/Halting_problem},
  2024.
\newblock [Online; accessed 3-July-2024].

\bibitem{om}
``{Oracle Machines}.'' \url{https://en.wikipedia.org/wiki/Oracle_machine},
  2024.
\newblock [Online; accessed 3-July-2024].

\bibitem{tcomp}
``{Turing Completeness}.''
  \url{https://en.wikipedia.org/wiki/Turing_completeness}, 2024.
\newblock [Online; accessed 3-July-2024].

\bibitem{cfg}
``{Control Flow Graph}.''
  \url{https://en.wikipedia.org/wiki/Control-flow_graph}, 2024.
\newblock [Online; accessed 3-July-2024].

\bibitem{gpath}
``{Path in Graph Theory }.''
  \url{https://en.wikipedia.org/wiki/Path_(graph_theory)}, 2024.
\newblock [Online; accessed 3-July-2024].

\bibitem{bva}
``{Boundary Value Analysis}.''
  \url{https://en.wikipedia.org/wiki/Boundary-value_analysis}, 2024.
\newblock [Online; accessed 3-July-2024].

\bibitem{eqc}
``{Equivalent Classes}.''
  \url{https://en.wikipedia.org/wiki/Equivalence_class}, 2024.
\newblock [Online; accessed 3-July-2024].

\bibitem{gbt}
``{Gray Box Testing}.'' \url{https://en.wikipedia.org/wiki/Gray-box_testing},
  2024.
\newblock [Online; accessed 3-July-2024].

\bibitem{eqcp}
``{Equivalent Partitioning}.''
  \url{https://en.wikipedia.org/wiki/Equivalence_partitioning}, 2024.
\newblock [Online; accessed 3-July-2024].

\bibitem{bigo}
``{Big Oh Notation}.'' \url{https://en.wikipedia.org/wiki/Big_O_notation},
  2024.
\newblock [Online; accessed 3-July-2024].

\bibitem{coup}
``{Software Coupling}.''
  \url{https://en.wikipedia.org/wiki/Coupling_(computer_programming)}, 2024.
\newblock [Online; accessed 3-July-2024].

\bibitem{jac}
``{Jaccard Index}.'' \url{https://en.wikipedia.org/wiki/Jaccard_index}, 2024.
\newblock [Online; accessed 3-July-2024].

\bibitem{dry}
``{Do Not Repeat Yourself}.''
  \url{https://en.wikipedia.org/wiki/Don\%27t_repeat_yourself}, 2024.
\newblock [Online; accessed 3-July-2024].

\bibitem{inc}
``{Incompleteness Theorems}.''
  \url{https://en.wikipedia.org/wiki/Gödel\%27s_incompleteness_theorems},
  2024.
\newblock [Online; accessed 3-July-2024].

\bibitem{cat}
``{Cat Source }.''
  \url{https://github.com/coreutils/coreutils/blob/master/src/cat.c}, 2024.
\newblock [Online; accessed 3-July-2024].

\bibitem{kc}
``{Kolmogorov Complexity }.''
  \url{https://en.wikipedia.org/wiki/Kolmogorov_complexity}, 2024.
\newblock [Online; accessed 3-July-2024].

\bibitem{bibo}
``{BIBO Stability}.'' \url{https://en.wikipedia.org/wiki/BIBO_stability}, 2024.
\newblock [Online; accessed 3-July-2024].

\bibitem{dyn}
``{Dynamical System}.'' \url{https://en.wikipedia.org/wiki/Dynamical_system},
  2024.
\newblock [Online; accessed 3-July-2024].

\bibitem{ms}
``{Metric Space}.'' \url{https://en.wikipedia.org/wiki/Metric_space}, 2024.
\newblock [Online; accessed 3-July-2024].

\bibitem{wchaos}
``{Chaos Theory}.'' \url{https://en.wikipedia.org/wiki/Chaos_theory}, 2024.
\newblock [Online; accessed 3-July-2024].

\bibitem{chaos1}
``{Definition of Chaos }.''
  \url{https://farside.ph.utexas.edu/teaching/329/lectures/node57.html}, 2024.
\newblock [Online; accessed 3-July-2024].

\bibitem{chaos2}
``{Characteristics of Chaos}.''
  \url{https://math.libretexts.org/Bookshelves/Scientific_Computing_Simulations_and_Modeling/Introduction_to_the_Modeling_and_Analysis_of_Complex_Systems_(Sayama)/09\%3A_Chaos/9.02\%3A_Characteristics_of_Chaos},
  2024.
\newblock [Online; accessed 3-July-2024].

\bibitem{le}
``{Lyapunov Exponent }.''
  \url{https://en.wikipedia.org/wiki/Lyapunov_exponent}, 2024.
\newblock [Online; accessed 3-July-2024].

\bibitem{sdchaos}
``{Software Development and Chaos Theory}.''
  \url{https://timross.wordpress.com/2010/01/17/software-development-and-chaos-theory/},
  2010.
\newblock [Online; accessed 3-July-2024].

\bibitem{KANTZ199477}
H.~Kantz, ``A robust method to estimate the maximal lyapunov exponent of a time
  series,'' {\em Physics Letters A}, vol.~185, no.~1, pp.~77--87, 1994.

\bibitem{fup}
``{Fourier Uncertainty Principle}.''
  \url{https://en.wikipedia.org/wiki/Fourier_transform#Uncertainty_principle},
  2024.
\newblock [Online; accessed 3-July-2024].

\bibitem{idm}
``{Iterative Development }.''
  \url{https://en.wikipedia.org/wiki/Iterative_and_incremental_development},
  2024.
\newblock [Online; accessed 3-July-2024].

\bibitem{pdm}
``{Software Prototyping}.''
  \url{https://en.wikipedia.org/wiki/Software_prototyping}, 2024.
\newblock [Online; accessed 3-July-2024].

\bibitem{Bakhtiary2020-yg}
V.~Bakhtiary, T.~J. Gandomani, and A.~Salajegheh, ``The effectiveness of
  test-driven development approach on software projects: A multi-case study,''
  {\em Bull. Electr. Eng. Inform.}, vol.~9, pp.~2030--2037, Oct. 2020.

\bibitem{tdd_3}
I.~Karac and B.~Turhan, ``What do we (really) know about test-driven
  development?,'' {\em IEEE Software}, vol.~35, pp.~81--85, 07 2018.

\bibitem{ccsc}
R.~P. Feynman, ``{Cargo Cult Science }.''
  \url{https://calteches.library.caltech.edu/51/2/CargoCult.htm}, 1974.
\newblock [Online; accessed 3-July-2024].

\end{thebibliography}
\bibliographystyle{ieeetr}

\end{document}